# Equilibrium Contact Angle at the Wetted Substrate

Leonid Pekker[a)], FujiFilm Dimatix, Inc. Lebanon, New Hampshire 03766, USA
David Pekker, University of Pittsburgh, Pittsburgh, Pennsylvania, USA
Nikolai Petviashvili, Independent Consultant, Inc. San Jose, CA 95119, USA

[a)]Author to whom correspondence should be addressed: leonid.pekker@fujifilm.com

**Abstract**

We construct a novel model for the steady-state contact angles of liquid droplets at the wetted substrate. The non-removable, thin liquid film covering the substrate is governed by the intermolecular forces between molecules of liquid and solid, which we describe using the standard disjoining pressure approximation. Balancing the disjoining pressure against the surface tension we find the smooth shape of the surface of the liquid. We show that we can extract an effective contact angle from the region where the film and the droplet meet. Crucially, we find that for large droplets the contact angle is independent of the droplet size. Instead, the contact angle is determined by the surface tension and the disjoining pressure parameters through a simple formula that works for both small and large contact angles. We suggest that comparing predictions of our model to experimentally measured contact angles will enable constraining the parameters of the disjoining pressure models.

**Introduction**

The wetting properties between the liquid and a solid substrate is determined by the cohesive interaction holding liquid molecules together, and the adhesive interactions between the liquid and solid molecules [1, 2]. The cohesive liquid interaction is described by $\gamma$, the surface tension coefficient, and the liquid-solid interaction by the Lennard-Jones type potential with a short-range repulsion term and a long-range decaying attraction term. Directly calculating the wetting properties of the liquid, like the contact angle between liquid and the substrate, based on Lennard-Jones potential is extremely difficult [1, 3, 4] and is not practical, see review [5]. However, the net effect of intermolecular potential on the wetting properties of a liquid film of thickness $h$ can be described by the disjoining pressure $\Pi(h)$, the net force per unit area of the liquid-solid interface [6-8]. For the case of horizontal substrate, $\Pi(h)$ is like the net gravitation pressure $\rho g h$. Various approximations for $\Pi(h)$ are used to model the nucleation and growth of dry zones in the process of dewetting of a solid surface covered by a thin liquid film [9-22], and to describe the physics of contact line motion and related problems [23-36].

In this paper, we focus on constructing a model for calculating the equilibrium (steady state) contact angles of droplets on wetted substrates covered by non-removable precursor thin liquid films. In our model, this type of films and their thicknesses are considered in the framework of disjoining pressure approximation. In the model, we use disjoining pressure in the form of Ref. [15]. We show that the contact angle increases with volume of the droplet for small droplets but saturates at a constant value for droplets with height greater than approximately five time the equilibrium thickness of the film. We would like to mention work [7] in which the author obtained the formula for small contact angles from different consideration; his formula is a special case of our model.

## II. Model of steady-state contact angle at wetted substrates

Let us consider the steady-state shape of a droplet placed on a substrate that supports a non-removable thin liquid film. Neglecting the gravitation, the equation describing the shape of the droplet in the cylindrical symmetrical case, Fig. 2, can be written as

$$\frac{d}{dZ}\left(-\frac{\gamma}{R_{\text{curv}}} - \chi\left\{\left(\frac{h^*}{h}\right)^m - \left(\frac{h^*}{h}\right)^n\right\}\right) = 0, \tag{1}$$

where

$$R_{\text{curv}} = -\frac{\gamma \frac{\partial^2 h}{\partial z^2}}{\left(1+\left(\frac{\partial h}{\partial z}\right)^2\right)^{1.5}}. \tag{2}$$

Eq. (1) states that the pressure at the free surface is balanced by the surface tension $\gamma$ associated with the surface curvature $R_{\text{curv}}$, Eq. (2), and the disjoining pressure of a film of thickness $h$ associated with intermolecular force model parameters $\chi$, $h^*$, $m$, and $n$ of Ref. [15] assuming $m > n$. In the model, we assume that the droplet is axisymmetric along the y-axis, Fig. 2.

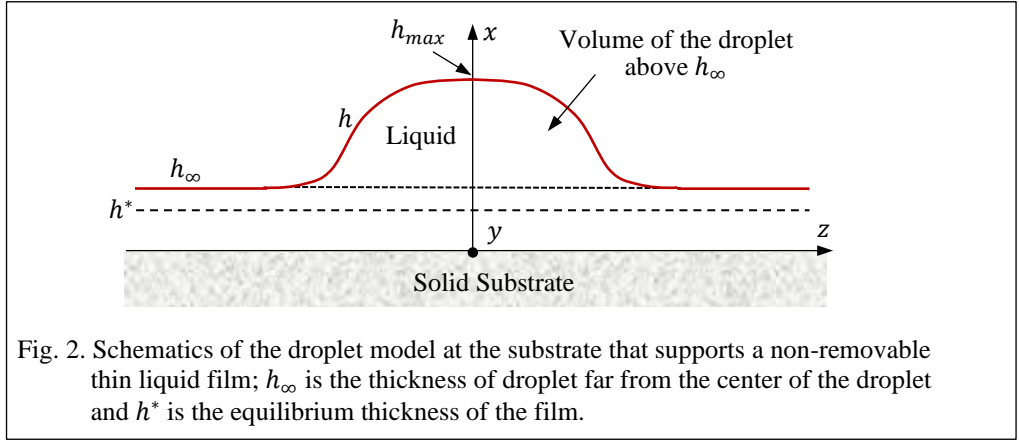

Fig. 2. Schematics of the droplet model at the substrate that supports a non-removable thin liquid film; $h_\infty$ is the thickness of droplet far from the center of the droplet and $h^*$ is the equilibrium thickness of the film.

Introducing the dimensionless variables $Z$, $H$, and the dimensionless lubrication parameter $\alpha_l$ defined as:

$$z = \sqrt{\frac{\gamma h^*}{\chi}} Z \quad h = h^* H \quad \alpha_l = \frac{\chi h^*}{\gamma}, \tag{3}$$

Eq. (1) becomes

$$\frac{d}{dZ}\left(\frac{\frac{d^2 H}{dZ^2}}{\left(1+\alpha_l\left(\frac{dH}{dZ}\right)^2\right)^{1.5}} + \frac{1}{H^m} - \frac{1}{H^n}\right) = 0, \tag{4}$$

and Fig. 1 becomes Fig. 3.

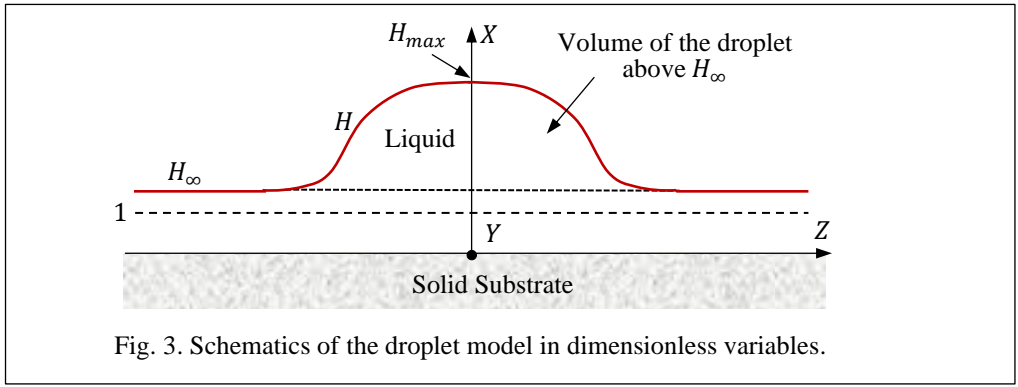

Fig. 3. Schematics of the droplet model in dimensionless variables.

We will now proceed to show that Eq. (4) admits a series of solutions that allow us to define an effective contact angle. Further, we show that the resulting contact angle only depends on $\alpha_l$, $m$, and $n$, but crucially is independent of the droplet volume.

Integrating Eq. (4) yields

$$\frac{\frac{d^2H}{dZ^2}}{\left(1+\alpha_l\left(\frac{dH}{dZ}\right)^2\right)^{1.5}} + \frac{1}{H^m} - \frac{1}{H^n} = -P, \tag{5}$$

where

$$P = \frac{1}{H_\infty^n} - \frac{1}{H_\infty^m} \tag{6}$$

is the pressure in the droplet and $H_\infty$ is the height of the film far away from the droplet (see Fig. 2). We observe that Eq. (5) just states that the pressure in the droplet is constant and independent of Z. We now consider two limits. For a droplet with $H(Z) \gg 1$ and sufficiently far from the droplet edges, we can neglect the effect of the intermolecular forces and this equation reduces to that of an ellipse in terms of the dimensionless variables $H$ and $Z$ as defined by Eq. (3). Eliminating the stretching of H relative to Z introduced by Eq. (3), we see that Eq. (4) is actually that of a circle, with the radius set by the pressure. On the other hand, far away from the droplet $Z \to \pm\infty$, we can drop the first term in the LHS of Eq. (5) and hence $H(Z \to \pm\infty) = H_\infty$. As we have already stated, the pressure is independent of the position and hence it must be balanced by the intermolecular forces which sets the value of $H_\infty$.

We will now establish a relation between the maximum height of a droplet $H_{max}$ and $H_\infty$ for large droplets $H_{max} \gg H_\infty$ (see Fig. 2). We proceed by multiplying Eq. (5) by $\frac{dH}{dZ}$ and integrating the resulting equation to obtain:

$$\left(\frac{dH}{dZ}\right)^2 = \frac{B-\alpha_l(0.5B)^2}{(1-0.5\alpha_l B)^2}, \tag{7}$$

$$B = 2\left(H\left(\frac{1}{H_\infty^m} - \frac{1}{H_\infty^n}\right) + \frac{1}{(m-1)H^{m-1}} - \frac{1}{(n-1)H^{n-1}} - \frac{m}{(m-1)H_\infty^{m-1}} + \frac{n}{(n-1)H_\infty^{n-1}}\right), \tag{8}$$

where we have used $H(Z \to \pm\infty) = H_\infty$ to set the constant of integration.

Next, let us determine the allowed values of $H_\infty$. First, the pressure in the droplet is positive and therefore $H_\infty > 1$. Second, we observe that the RHS of Eq. (6) reaches its maximum value when $H_\infty = \sqrt[m-n]{m/n}$, which sets the upper bound on $H_\infty$. In summary: $1 < H_\infty < \sqrt[m-n]{m/n}$.

We now focus on large droplets for which $H_{max} \gg H_\infty$. From Eq. (7) it follows that $H_{max}$ correspond to $B = 0$ (since $\frac{dH}{dZ} = 0$ at the top point of the droplet). Further, as pressure is low in large droplets, we can write $H_\infty = 1 + \varepsilon$, where $\varepsilon \ll 1$. Substituting this expression into Eq. (8) and dropping the small terms $\frac{1}{(m-1)H_{\max}^{m-1}} - \frac{1}{(n-1)H_{\max}^{n-1}}$ we obtain a relation between $H_{\max}$ and $H_\infty$:

$$H_{\max} = \frac{1}{\varepsilon(m-1)(n-1)} = \frac{1}{(H_\infty-1)(m-1)(n-1)}. \tag{9}$$

We observe that as $H_{\max}$ increases (and so does the droplet volume), the height at infinity $H_\infty$ decreases approaching unity for very large droplets. This relation is consistent with the fact that the pressure due to surface tension in large droplets is smaller than the pressure in smaller droplets.

Our next step is to determine the effective contact angle $\theta_c$. Specifically, we compute $\tan\theta_c = \sqrt{\alpha_l}\frac{dH}{dZ}$ at the inflection point $H_i$ (defined by $\left.\frac{d^2H}{dZ^2}\right|_{H_i} = 0$), where the factor of $\sqrt{\alpha_l}$ compensates for the stretching of H relative to Z introduced by the transformation to dimensionless coordinates in Eq. (3). Substituting

$$H_\infty = 1 + \varepsilon \tag{10}$$

into Eq. (5), and using $\varepsilon \ll 1$, we obtain that $H$ at the inflection point, $H_i$, satisfies the equation

$$\frac{1}{H_i^n} - \frac{1}{H_i^m} = (m-n)\,\varepsilon. \tag{11}$$

Further, at the inflection point we expect that $H_i \gg 1$, and therefore we can drop the second term on the LHS of Eq. (11) to obtain the expression

$$H_i = \frac{1}{\sqrt[n]{(m-n)\varepsilon}}. \tag{12}$$

Comparing Eqs. (9) and (12), we note that

$$\frac{H_i}{H_{\max}} = \left(\varepsilon^{\frac{n-1}{n}}\right)\frac{(m-1)(n-1)}{\sqrt[n]{(m-n)}} \ll 1. \tag{13}$$

Substituting Eq. (12) into Eq. (8), and again taking into account that $\varepsilon \ll 1$, we obtain $B(H_i) = \frac{2(m-n)}{(m-1)(n-1)}$.

Finally, substituting $B(H_i)$ into Eq. (7), we obtain a formula for contact

$$\tan(\theta_c) = \left(\frac{\chi h^*}{\gamma}\right)^{1/2} \frac{\left(\frac{2(m-n)}{(m-1)(n-1)} - \frac{\chi h^*}{\gamma}\left(\frac{(m-n)}{(m-1)(n-1)}\right)^2\right)^{1/2}}{1 - \frac{\chi h^*}{\gamma}\frac{(m-n)}{(m-1)(n-1)}}, \tag{14}$$

where we have substituted the value of $\alpha_l$ from Eq. (3).

Before discussing the implications of Eq. (14), we present numerical calculations that support the validity of the approximations that we made in deriving Eq. (14). Specifically, we compare the droplet shapes obtained by Eqs. (7, 8) and the effective contact angle model implied by Eq. (14). The effective contact angle model combines a circular-cap droplet shape, with the radius of curvature set by the droplet pressure given by Eq. (6) with the constraint on the contact angle (which is applied at the height $H = 1$) set by the effective contact angle formula. We expect that the droplet shapes obtained using these two models should coincide, except near the substrate for $H \lesssim H_i$, where the numerical solution of Eqs. (7, 8) should develop a "foot" like feature which is the result of the action of the inter-molecular forces. We plot the droplet shapes obtained using the two different models in Fig. 3. To construct the plots, we consider both small contact angles $\alpha_l = 0.3$ (Fig. 1a) and large contact angles $\alpha_l = 5.0$ (Fig. 1b). We also sweep through a range of values of $\varepsilon$, which is related to the pressure through Eqs. (6) and (10) and hence to the droplet size. Finally, we set the inter-molecular force exponents to $m = 9$ and $n = 3$ [2]. We note that in order to have a natural aspect ratio between $H$ and $Z$ axes in Fig. 1, we needed to shrink the $Z$ axes by the factor $\sqrt{\alpha_l}$ to compensate for the corresponding factor in the definition of dimensionless variables in Eq. (3). In Fig. 3, we observe that, except for the aforementioned foot feature, the two models predict very similar droplet shapes down to remarkably small droplet sizes. The circular cap of the droplets predicted by the two models is essentially indistinguishable until the droplet height becomes only about 5 times the thickness of the non-removable film ($H_{\max} \sim 5$ which corresponds to $\epsilon \sim 0.02$). This observation holds for both small contact angles (see Fig. 1a) as well as for large contact angle $\theta_c > \pi/2$ (when $\frac{\chi h^*}{\gamma} \frac{(m-n)}{(m-1)(n-1)} > 1$) (see Fig. 1b), thus validating Eq. (14).

We now comment on the significance of Eq. (14). As one can see, Eq. (14) for the contact angle is independent of the volume of the droplets (volume of droplets above $h_\infty$, Fig. 1). Moreover, since $h_i$, the droplet height at the inflection point where we calculate the contact angle, is much larger than the equilibrium thickness of the film $h^*$, $\frac{h_i}{h^*} \sim \frac{1}{\sqrt[n]{\varepsilon}}$ (see Eq. (12)), and at the same time is much smaller than

$h_{max}$, $\frac{h_i}{h_{\max}} \sim \varepsilon^{\frac{n-1}{n}}$ (see Eq. (13)), we argue that this expression is the true steady-state (equilibrium) contact angle for non-microscopic droplets ($\varepsilon \ll 1$), i.e. droplets with linear size $h \gg h^*$. We also remark that for the case of $\frac{\chi h^*}{\gamma} \ll 1$, Eq. (14) reduces to expression [7]:

$$\theta_c = \sqrt{\frac{2\chi h^*(m-n)}{\gamma(m-1)(n-1)}} \ . \tag{15}$$

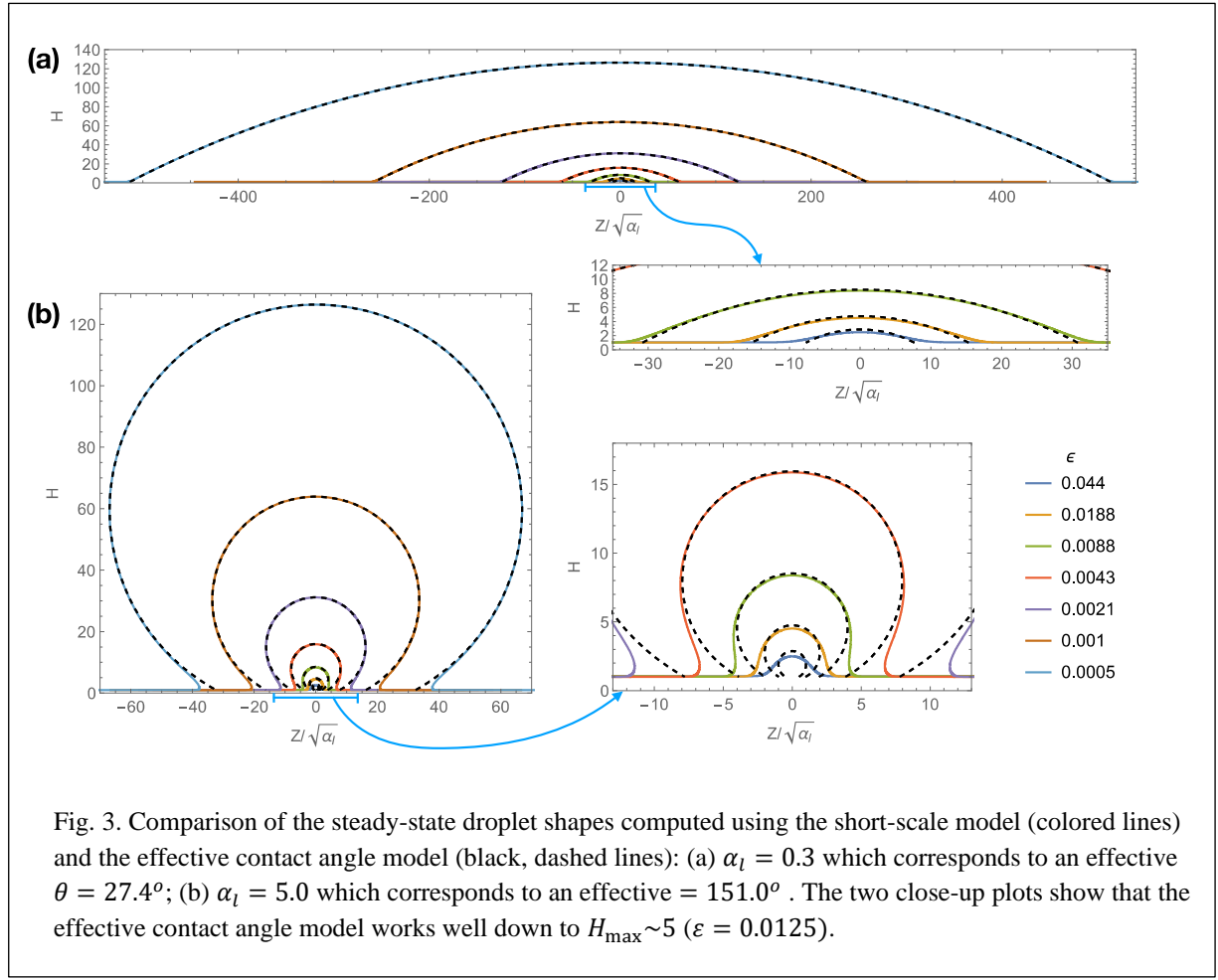

Fig. 3. Comparison of the steady-state droplet shapes computed using the short-scale model (colored lines) and the effective contact angle model (black, dashed lines): (a) $\alpha_l = 0.3$ which corresponds to an effective $\theta = 27.4°$; (b) $\alpha_l = 5.0$ which corresponds to an effective $= 151.0°$. The two close-up plots show that the effective contact angle model works well down to $H_{\max} \sim 5$ ($\varepsilon = 0.0125$).

### III. Concluding remarks

In this article, we obtained a novel formula for the steady-state contact angle at the wetted substrate covered by precursor non-removable thin liquid films. In the model, the films are considered in the disjoining pressure approximation. The obtained formula works for the full range of contact angles and agrees with the classic formula of Ref. [7] obtained for the case of small contact angles. Remarkably, our numerical testing shows that our formula works well not only for large droplets where the heights of droplets is much larger than the equilibrium thickness of the film, $H_{\max} \gg h^*$, but also for droplets with heights down to $5h^*$.

Finally, we remark that the disjoining pressure model parameters can be experimentally constrained by comparing the experimentally observed contact angle data against Eq. (14).

**ACNOLEDGEMENTS**

The authors would like to express their sincere gratitude to Dan Barnett for his kind support of this project and to Paul Hoisington, Chris Menzel, James Myrick, and Matthew Aubrey for helpful discussions.

**AUTHORS DECLARATIONS**

**Conflict of interest**

Authors have no conflict to disclose

**DATA AVAILABILITY**

The data that support the findings of this study are available from the corresponding author upon reasonable request.